# Multivariate Microaggregation of Set-Valued Data

**Malik Imran-Daud**

Department of Software Engineering, Foundation University Islamabad; e-mail: imrandaud@fui.edu.pk

**Muhammad Shaheen**

Department of Software Engineering, Foundation University Islamabad; e-mail: dr.shaheen@fui.edu.pk

**Abbas Ahmed**

Department of Electrical Engineering, Beaconhouse International College, Islamabad;
e-mail: aa28g08@southamptonalumni.ac.uk

**Corresponding author:** imrandaud@fui.edu.pk

Data controllers manage immense data, and occasionally, it is released publically to help the researchers to conduct their studies. However, this publically shared data may hold personally identifiable information (PII) that can be collected to re-identify a person. Therefore, an effective anonymization mechanism is required to anonymize such data before it is released publically. Microaggregation is one of the Statistical Disclosure Control (SDC) methods that are widely used by many researchers. This method adapts the k-anonymity principle to generate *k*-indistinguishable records in the same clusters to preserve the privacy of the individuals. However, in these methods, the size of the clusters is fixed (i.e., *k* records), and the clusters generated through these methods may hold non-homogeneous records. By considering these issues, we propose an adaptive size clustering technique that aggregates homogeneous records in similar clusters, and the size of the clusters is determined after the semantic analysis of the records. To achieve this, we extend the MDAV microaggregation algorithm to semantically analyze the unstructured records by relying on the taxonomic databases (i.e., WordNet), and then aggregating them in homogeneous clusters. Furthermore, we propose a distance measure that determines the extent to which the records differ from each other, and based on this, homogeneous adaptive clusters are constructed. In experiments, we measured the cohesiveness of the clusters in order to gauge the homogeneity of records. In addition, a method is proposed to measure information loss caused by the redaction method. In experiments, the results show that the proposed mechanism outperforms the existing state-of-the-art solutions.

## 1. Introduction

Due to rapid developments in information technology, the users of such technologies are growing day by day. As a result, these technologies generate a huge amount of data in the form of emails, query logs, social network activities, medical records, etc (known as set-valued data). A set-value data contains a set of elements associated with a specific person that is extracted from the dataset (e.g., name, identity, disease, etc). Hence, online data management is turning out to be a real concern for data controllers (e.g., search engines, hospitals, organizations, etc.). The data controllers occasionally release such data (i) to help the researchers in order to improve their methods, (ii) to assess their theories or hypotheses [16, 17], and (iii) even it is released to the marketing companies to find the relevant people to promote their products.

The publically shared data may hold personally identifiable information (PII) of individuals, which can be collected and analyzed to re-identify a person [18]. Therefore, such data can be exploited for unfair purposes, hence, it raises disclosure or privacy risks for the individuals [44]. Usually, the datasets that hold PII comprise of the following type of key variables [45, 48]: (i) identifying variables (value of such variables identifies a person e.g., name, social security number, age, date of birth, etc.), and (ii) sensitive variables (as defined by the legislation [15, 19], values of such variables may cause legal implications e.g., political views, religious views, diseases, etc). The information collected from a set of identifying variables can also be linked together to re-identify an individual, such variables are known as quasi-identifiers (QIs). For example, a combination of variables like age, gender, and location can re-identify a specific person. Thus, publically shared data must be de-identified before it is released publically in order to ensure the privacy of the individuals.

In most of the cases, the data being used by the applications or a specific environment is managed in structured (e.g., database records organized in a tabular format) or unstructured format (e.g., web search query logs, social network data, medical reports, etc.). In structured data, the data analysis is simple, however, it is a challenging job in unstructured data (i.e., set-valued data) due to its volume and diversity in the semantics [13]. Hence, the de-identification process requires a rigorous analysis of the data in order to elicit such variables (i.e., identifying variables and sensitive variables).

To achieve the aforementioned objectives, the Statistical Disclosure Control (SDC) methods [48] are proposed to protect data from disclosure risks by anonymizing PII. Researchers [16, 22, 43, 44, 51] extended these methods to further improve the redaction techniques. Furthermore, there are a few more methods that are to add noise to the data [11, 37], microaggregation [9, 43], and others [48]. Microaggregation is a well-known SDC method that has attracted researchers' attention. This method incorporates the principle of k-anonymity [40, 47]. K-anonymity principle minimizes disclosure risks by suppressing or generalizing key variables (i.e., identifying or sensitive variables), such that k records exhibit similar characteristics and they are indistinguishable. However, an adversary aware of a person's background may re-identify an individual from k-anonymized records that hold similar values. To solve this problem, the principle of $l$-diversity [25] suggests that key variables in records should have diverse $l$-values. Although, the $l$-diversity is prone to skewness and similarity attacks [24], hence, the t-closeness [24] principle further enhances the privacy of the $l$-diversity principle by managing the distribution of the data values of the attributes. This principle suggests that the distance between the sensitive attributes in the equivalence class and the entire table should not exceed the $t$-threshold.

Researchers have proposed several microaggregation-based tools/techniques [46]. The microaggregation based methods aggregate homogeneous records in common clusters that exhibit similar characteristics of data (i.e., each cluster holds at least $k$ similar records). These records are then replaced by the centroid (a record) of the cluster to ensure that the records are indistinguishable (so-called record anonymization). However, in such methods, the utility of de-identified data has always been a serious concern. In addition, these approaches bear some limitations, which are (i) the dimensionality issues [2] (i.e., greater loss of information on large datasets), and (ii) it is difficult to identify sensitive variables from the unstructured data [13]. Moreover, the privacy issues raised by the $l$-diversity and $t$-closeness principles also require special attention. A few of the approaches based on such principles are discussed in Section 2.

MDAV [10] is a well-known microaggregation-based algorithm that partitions the dataset into homogeneous clusters. This algorithm requires the following to operate (i) a centroid of the dataset, and (ii) a distance measure that computes the distance between the records to assess their similarity. Consequently, the records are aggregated in clusters based on their distance scores. Initially, the MDAV was proposed for numerical datasets that map the univariate in clusters. Later, it was adapted for the categorical datasets by various researchers [1, 5, 28]. However, the clusters generated through these methods are less cohesive; as a result, the utility of the anonymized data is also reduced. In addition, the records of the clusters can hold some uncommon attributes (e.g., the attributes present in one record, but missing in another record); hence, the distance that is measured based on such attributes may generate incorrect results.

### 1.1. Our Contributions

By considering the limitations discussed in Section 1, in this paper, we present a novel approach to anonymize unstructured set-valued data. Following are the novel contributions:

- In this paper, we extend our work [21] and improve the existing approaches [5, 28] with the following: (i) we propose an extension to the MDAV microaggregation method [10] to support the unstructured categorical data (i.e., improving structured approaches [21, 28]), and (ii) we improve the distance measure to consider semantically similar and non-similar attributes as separate variables within the tuples while computing distance between the queries (in contrast, [5] treats them as a single record that may result in information loss). Semantically similar attributes are the attributes that have the same semantics that is determined through the lexical database from the distance measure (i.e., the matching attributes have the least semantic distance), whereas, non-semantically similar attributes do not have any matching attribute with the same semantics (or the taxonomic distance is higher). The cardinality of semantically similar and non-similar multivariate in distinct records are used to assess the level of contrast between the matching tuples. In addition, the newly proposed semantic distance-based method relies on (i) the interrelationship of taxonomic levels of the attributes (for example, under the sports category, the distance scores are different when we compare a football with the soccer and a football with the circle), and (ii) the cardinality of the attributes of the distinct records.
- The proposed solution relies on an adaptive clustering approach (i.e., the size of clusters is not fixed). The records are aggregated in similar clusters based on their semantic similarity score. The clusters generated as a result of this process are (i) more cohesive, and (ii) they emit less information loss as compared to the fixed and variable size clusters [28], which is later proved through the experiments.
- In addition, we derive metrics to measure cohesiveness and information loss of the clusters in order to evaluate the effectiveness of our proposed model.

The rest of the paper has the following content. Section 2 illustrates the state-of-the-art solutions related to anonymization. In Section 3, we present our proposed scheme to anonymize set-valued data. Moreover, we evaluate our proposed scheme in Section 4. Section 5 illustrates the conclusion and future work.

## 2. Related Work

Researchers have significantly contributed to de-identify a person by sanitizing the sensitive information published in the form of text. For this purpose, the symmetrical records are grouped in similar clusters based on their common properties. To achieve this, the researchers have proposed several clustering techniques [35, 36, 39, 50] that operate in diverse environments (e.g., big data, cloud, etc.). Chakaravarthy et al. [8] proposed a privacy model to protect the information published over the internet. In this approach, it is assumed that the context terms are associated with the entities, and an attacker can re-identify a person through these associated sensitive terms. Therefore, a document is sanitized by masking the context terms that can be used by an adversary to re-identify a person. This concept was adopted by Anandan et al.

[4], hence, they proposed a model that relies on the taxonomic interrelations of the sensitive terms that are identified manually. In this model, at least t-plausible documents are generated by sanitizing the sensitive terms through the generalized entities derived from the taxonomy. This model helps to generalize the sensitive terms of a single record, but it requires improvements to quantify the similarity between the multiple records in order to fulfill the requirements of microaggregation.

In addition, the research community has also proposed several solutions to anonymize structured and unstructured data, few of them are discussed in accordance with the limitations highlighted in the previous section (i.e., Section 1). First, we discuss some approaches that are used to anonymize structured data. Domingo-Ferrer and Soria-Comas [13] examined the possible ways to use k-anonymity model in the big data environment and, based on this, they proposed a solution that enhances the capabilities of this model by dealing with (i) the issues related to the dimensionality of the attributes and (ii) the issues of releasing overlapping anonymized data from the multiple sources. To accomplish these objectives, the proposed model suggests constructing multiple sets of QIs, and, based on these sets, various versions of the k-anonymized data are generated. These sets are ascertained to hold one element of sensitive attributes. By doing so, it hampers the possibility of appearing the maximum number of sensitive attributes in a single set of QIs; hence, this approach addresses the above-mentioned issues. However, this solution is constrained to structured data where sensitive attributes are already known or predetermined, and it is not suitable for the unstructured set-valued data. In another approach, Xu et al. [51] proposed a utility-based anonymization framework that relies on the local recording method [48]. This method is based on the quality metrics that measure the utility of numeric and categorical attributes by relying on the taxonomy. In this framework, the records are anonymized by replacing the value of an attribute with all possible leaf nodes of the taxonomic branches; consequently, it constructs multiple anonymized records for each given attribute. Hence, these records are arranged in multiple clusters based on their similarity, and then the weighted penalty of each cluster is measured. As a result, a cluster that has less weighted penalty is assumed to cause less information loss; as a result, the anonymized records of this cluster are released to the public. As this approach replaces the values of the records with the leaf nodes of the taxonomy, thus, it still causes information loss or even distorts the information. For example, Hepatitis and HIV are two different types of leaf nodes that correspond to the common ancestor node disease within the taxonomy. Thus, it distorts the information if all values of a dataset are replaced with any of these leaf nodes. In addition, it involves overhead to form clusters and then calculating their weighted penalty for the release purpose.

Likewise, Majeed et al. [26] proposed another framework that measures the identity vulnerability of the QIs and the diversity of the sensitive attributes (SAs) before anonymizing the structured data. To do so, this framework relies on the random forest method [7] to measure identity vulnerabilities in the data (an artificial intelligence-based technique to identify sensitive QIs that may reveal information). To achieve this objective, this method calculates the vulnerability scores of the QIs in order to highlight such QIs that require anonymization. Finally, the chosen QIs and SAs are anonymized with the generalized data items determined from the taxonomic tree. Again, this method relies on predetermined attributes (i.e., QIs and SAs) that may not be suitable in the unstructured set-valued data.

In addition, researchers have proposed several models to anonymize unstructured data. Motwani and Nabar [29] proposed a mechanism that optimizes the k-anonymity model to deal with the anonymity issues in the unstructured data (i.e., online query logs). For this purpose, they proposed several algorithms that achieve the notion of k-anonymity after transforming unstructured data into a structured format. The notion of k-anonymity is achieved by adding or deleting the supplementary data items into the records to make them indistinguishable. However, the fabrication of data items distorts the actual records; thus, it could lead to inaccurate results during the testing of the hypothesis. In another similar approach, Terrovitis et al. [49] proposed an anonymization mechanism inspired by the k-anonymity model for the unstructured attributes. To anonymize data, the proposed algorithms replace the semantically similar data items of the records with the generalized node of the taxonomic branch of the lexical database. Moreover, they proposed a metric to measure the information loss caused by such generalizations. This approach performs better than the previous approach (i.e., [29]) (in terms of data loss), as it anonymizes data items with the generalized node that is semantically similar and retains the semantics of attributes.

In another approach, Gardner and Xiong [16] proposed a framework to identify personally identifiable information from the unstructured data through the existing natural language processing techniques [30], which is later linked to the individuals before anonymization. This approach leverages the HIPAA [19] rules to consider the sensitive attributes that require protection, but it refrains to present any model that detects such information from the published data. In addition, this approach relies on the Mondrian multidimensional approach [23] (a k-anonymity based approach for multidimensional attributes) to anonymize sensitive information.

Differential privacy protection [14] is another model that relies on a concept to add a small amount of noise to the data (i.e., ε) to protect the data records against prediction. In a similar approach, Parra-Arnau et al. [34] proposed a microaggregation-based model that relies on the differential privacy method to limit disclosure risks. Moreover, several algorithms are proposed that microaggregate a complete record or a specific group of attributes within a record to preserve privacy. Sánchez et al. [45] proposed another differential privacy-based method to preserve the privacy of the query logs. In this approach, the idea was to preserve the semantics of the query logs through the differential privacy-based method by preserving the cardinality and granularity as the original query logs. Differential privacy is a better method than the k-anonymity method to protect privacy [45]. However, in this paper, we are improving the MDAV approach and, in the future, we will study the differential privacy aspect in the same model.

To summarize, the state of the art mainly focuses on the structured data to anonymize personally identifiable information (PII); however, the solutions that operate on the unstructured data are unable to provide a concrete solution that could identify PII automatically. Moreover, the anonymization methods distort information during the data generalization phase; as a result, they cause huge information loss. In contrast, we propose a microaggregation-based framework that automatically identifies sensitive information from the unstructured data, as defined by the legislation [15, 19], and then, de-identifies this information by generating at least k-similar records. To find similarities in the records, the system semantically analyzes such records by relying on the taxonomic lexical database (i.e., WordNet), and then aggregates them in homogeneous clusters. The similarity factors include both the semantically similar and non-similar attributes that are used to scale the level of similarity of the records (which most of the existing approaches lack). In addition, the size of the clusters is not fixed but adaptive in nature; as a result, the homogeneity of clusters is guaranteed. In addition, we provide a metric to measure the utility of the anonymized data that assures the proposed method emits less information loss.

## 3. Proposed Framework

As stated in Section 1, the primary objective of the proposed microaggregation-based technique is to construct such clusters that hold homogeneous records. Hence, in our proposal, the set-valued records (e.g., query logs) are classified in n-clusters, where each cluster holds at least k similar records to comply with the notion of k-anonymity principle. This adaptive nature of the clusters guarantees homogeneous records, thus, it ensures that the clusters are more cohesive (which is later proved through the experiments). To construct clusters, the mutual distance of the records is computed based on the similarity of their multivariate. Finally, these records are microaggregated through the extended MDAV algorithm, which replaces semantically similar records with the centroid of the given cluster. Figure 1 illustrates the framework to microaggregate set-valued data, which is discussed below:

- Semantic data analysis: As set-valued data is unstructured in nature, therefore, it is required to identify sensitive attributes from the records, which are then microaggregated in the later phase. For this purpose, the query logs are semantically analyzed in order to identify personally identifiable information (PII) from the records (PII are defined by the legislation [15, 19] and discussed in Section 1) (details in Section 3.1).

**Figure 1**
Anonymization of Set-valued Data

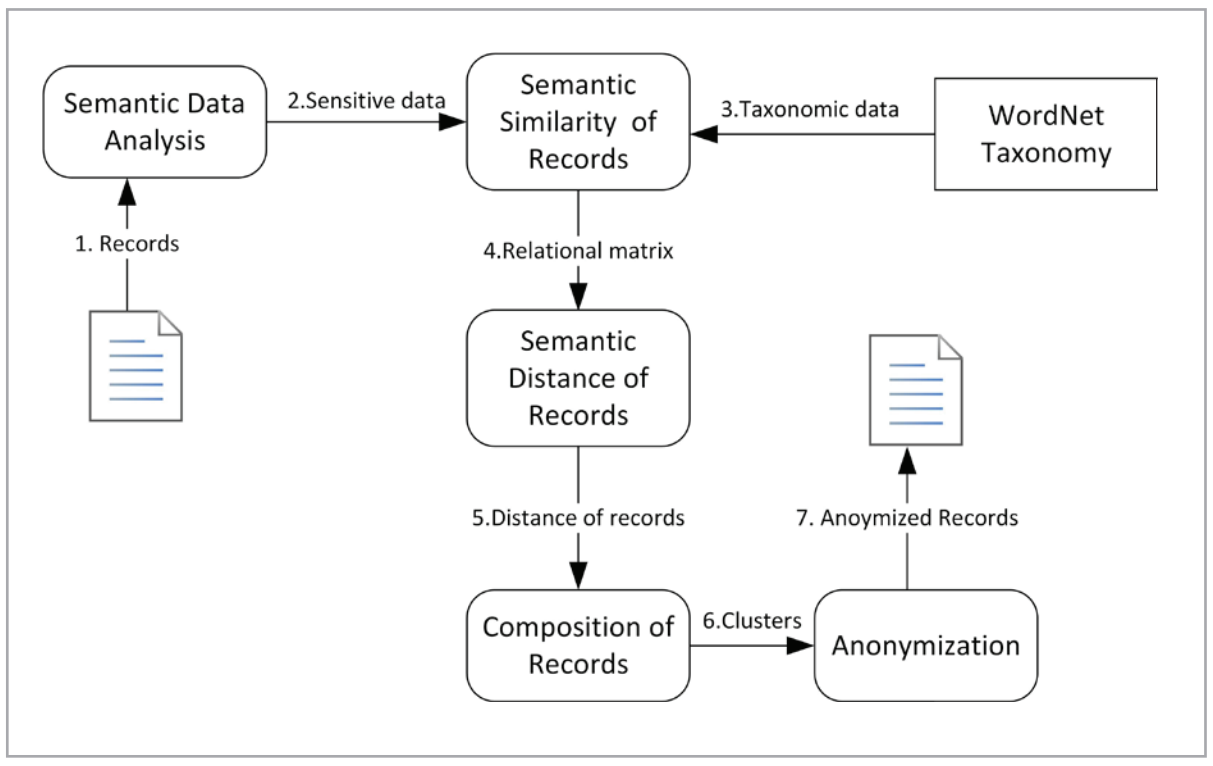

- Semantic similarity of the records: Once the sensitive data is retrieved, the records are then analyzed to identify multivariate that are semantically similar. For this purpose, we rely on WordNet taxonomy that holds the conceptual abstraction of the data elements. As a result of this step, a matrix is formed that holds distinct key-variables and their nature of association in multiple records (i.e., the extent to which they are similar) by highlighting semantically similar and non-similar attributes.
- Semantic distance and composition of records: In order to aggregate records in clusters, it is essential to compute the semantic distance of the records. For this purpose, the records are analyzed to compute the distance of each possible pair of records. In this comparison, it computes the maximum number of multivariate that are semantically similar in each pair of the records. As a result, an aggregated distance of each pair of records is computed, which is used to construct clusters comprising of homogeneous records (details in Section 3.2).
- Anonymization: Once the records are managed in their respective clusters, we compute the centroid of each cluster that replaces other records to make them k indistinguishable records (details in Section 3.3.

## 3.1. Semantic Analysis

As the query logs are unstructured in nature, therefore, a semantic analysis is required to identify multivariate that holds sensitive information. In our study, we consider only those attributes as sensitive that are defined by the legislation [15, 19] (e.g., topics like health, race/ethnic, religious views, political views, age, gender, name, etc). In the parts of speech, the nouns and verbs hold informative content that can be sensitive in nature [42], hence, the query logs are analyzed to identify sensitive nouns and verbs. For this purpose, we rely on a taxonomic database to identify such sensitive attributes. These sensitive attributes can be identified through the taxonomic branch of the sensitive topics (which are declared sensitive by the legislation). For example, the types of diseases like flu and pneumonia (as shown in Figure 2) fall under the category of Disease, which is declared sensitive by the legislative bodies. Hence, attributes like flu and pneumonia are considered sensitive.

**Definition 1:** A set of queries $Q = \{q_1, q_2, .....q_n\}$ be processed by the system to identify the sensitive information. Where, each query qn has a set of attributes $a_{1n}, a_{2n}, .....\ a_{mn}$ (i.e., nouns that hold sensitive content), where $a_{mn}$ is the $m^{th}$ attribute of the $n^{th}$ query. Thus, the query sets along with their attributes are listed below:

$$q_1 = \{a_{11}, a_{21}, a_{31} ......a_{i1}\}$$
$$q_2 = \{a_{12}, a_{22}, a_{32} ......a_{j2}\}$$
$$q_3 = \{a_{13}, a_{23}, a_{33} ......a_{k3}\}$$
$$q_n = \{a_{1n}, a_{2n}, a_{3n} ......a_{ln}\}$$

In order to identify noun phrases from the query records, we rely on our already proposed system [20] that illustrates the procedure to identify parts of speech through the natural language processing libraries [33]. After the data analysis step, we have a set of multivariate from each query record as listed in Definition 1. In the next step, we analyze the semantics of the query records in order to quantify the level of similarity they hold with each other. Based on the similarity score, the query logs are aggregated into homogeneous clusters. For this purpose, the queries are processed in the following order: (i) identify semantically similar and non-similar attributes within the query records, and (ii) based on distinct attributes determined in step (i), identify the semantic similarity between the pair of query records. The later step reduces computation time by matching only distinct attributes as determined in the former step (i.e., considering only one instance of all semantically similar attributes of the records). In order to find semantic similarity between attributes, we rely on a knowledge base (i.e., WordNet) that provides a conceptual representation of the attributes arranged in taxonomic order. To do so, we use a taxonomy-based measure [41] (shown in Equation (1)) that calculates the semantic distance between the given attributes (e.g., $var_1$ and $var_2$) from the taxonomic database. A semantic distance of the attributes illustrates the level of dissimilarity between their semantics. This measure illustrates the ratio between the distinct taxonomic ancestors of the variables (i.e., $T(var_1)$ and $T(var_2)$), divided by their cumulative ancestors. The logarithmic scale adds the smoothing factor between the variables, whereas the factor '1+' avoids the condition of log(0).

$$\delta(\text{var}_1, \text{var}_2) = \log_2\left(1 + \frac{|T(\text{var}_1) \cup T(\text{var}_2) - T(\text{var}_1) \cap T(\text{var}_2)|}{|T(\text{var}_1) \cup T(\text{var}_2)|}\right). \quad (1)$$

**Example 1:** To formulate the working of Equation (1), we compute the semantic distance (i.e., $\delta(var_1, var_2)$) between the attributes flu and pneumonia. For this purpose, we derived a taxonomic relation of these attributes (as illustrated in Figure 2) from the WordNet (a taxonomically arranged knowledgebase). In this taxonomy, these attributes have 12 distinct ancestors (i.e., $T(var_1) \cup T(var_2)$), out of which 10 are the common ancestors ($T(var_1) \cap T(var_2)$). As a result of Equation (1), the semantic distance between these attributes is 0.22, which states that these attributes are semantically coherent.

Based on Equation (1), the semantic similarity between two variables is determined by computing the inverse of semantic distance, which is represented by the following equation (i.e., Equation (2)). This similarity score ranges between '0' and '1', where a score close to '0' represents dissimilar attributes and a score closer to '1' signifies that the two attributes have a high similarity index and they are highly semantically similar attributes.

$$sim(\text{var}_1, \text{var}_2) = 1 - \delta(\text{var}_1, \text{var}_2). \quad (2)$$

To compute semantic similarity between the attributes, a query $q_n = \{a_{1n}, a_{2n}, \ldots\ldots a_{pn}\}$ holding a set of multivariate is processed by the system, and the similarity score of each distinct pair of attributes is computed through Equation (2). As a result, a pair of noun phrase that attains a threshold of similarity score (e.g., similarity > 0.8) is considered highly semantically similar attributes. To achieve this, Algorithm 1 illustrates a procedure to compute the similarity between each pair of the attributes of a query. In Algorithm 1, the similarity is calculated by matching each attribute of a query record $q_n$ with the rest of its attributes, and their distance is computed (using Equation (2)) based on the taxonomically arranged knowledgebase (i.e., WordNet) (lines 1-4). As a result, all pairs of the attributes that attain a threshold level of similarity are chosen for further processing (lines 5-10).

**Figure 2**

The taxonomic flow of attributes flu and pneumonia

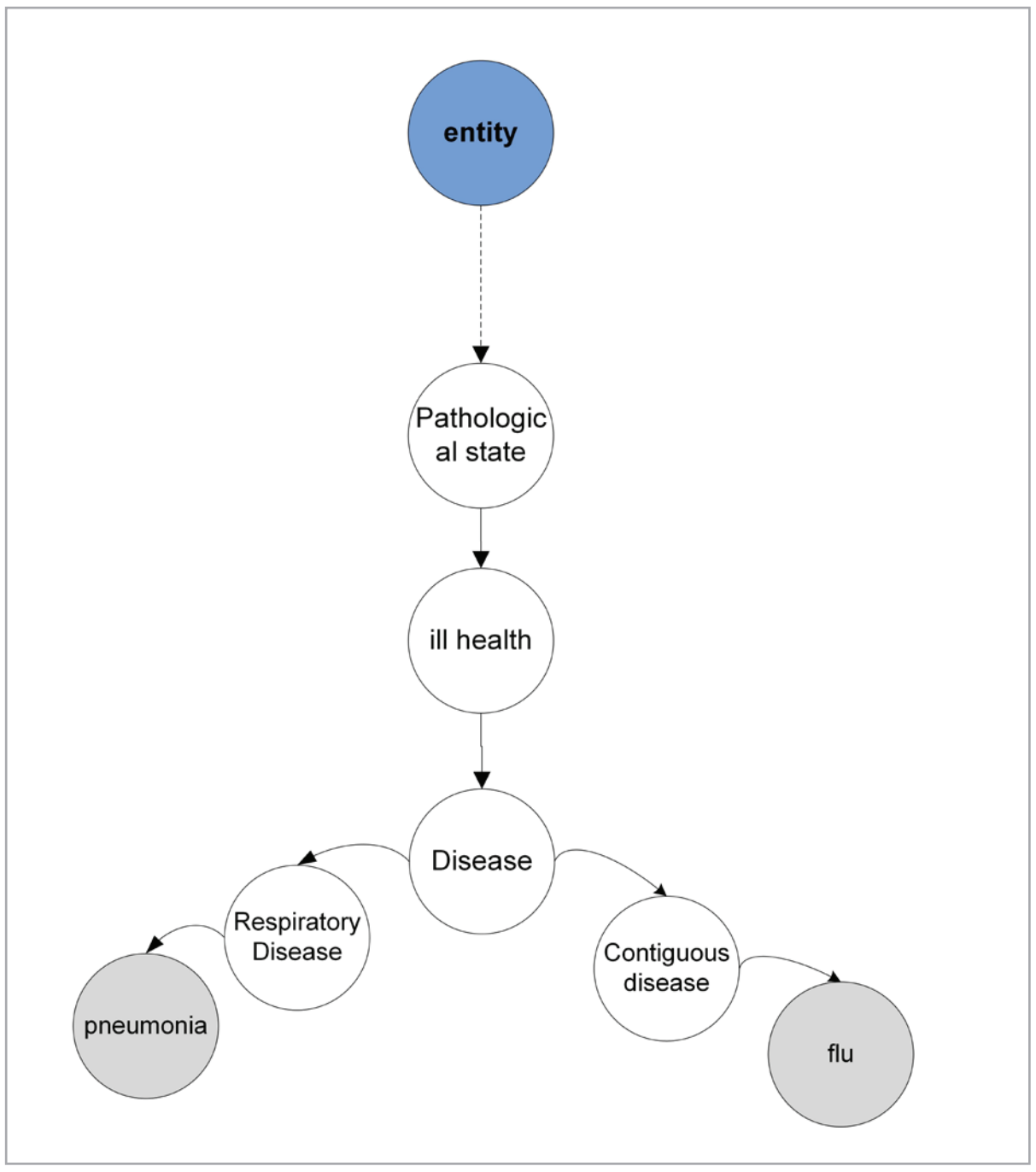

**Algorithm 1:** *SimilarityOfAttributes* ($q_n$)

Define array *similarity[n][n]*
**loop** starts at j← 1 till total_attributes - 1
  **loop** start at k←j+1 till total_attributes
    *similarity[j][k]* = (1- *TaxonomicDistance*(*attributes[j]*, *attributes[k]*))
    **if** *similarity[j][k]* >= max
      *Indexes* ← Save indexes of variables *j* and *k* as they both are similar
    **end if**
  **end loop**
**end loop**
**return** indexes

**Example 2:** A sample of a query record (e.g., a cure for ill-health pulmonary disease lung infection) is taken to demonstrate the working of the proposed solution. A query has a set of multivariate $q_1$={ $a_{11}$, $a_{21}$, $a_{31}$, $a_{41}$, $a_{51}$} (where $a_{11}$=infection, $a_{21}$=disease, $a_{31}$=ill-health, $a_{41}$=cure, $a_{51}$=lung). Algorithm 1 measures the semantic distance of each pair of the attributes of a query record (through the measures defined in Equation (1) & Equation (2)). Table 1 holds the similarity score of each distinct pair of the attributes of a query $q_1$. In Table 1, the attributes $a_{11}$, $a_{21}$, and $a_{31}$ (i.e., infection, disease, and ill-health respectively) are semantically similar as they hold higher similarity scores. Therefore, only one instance of such attributes is used in the next step (which is to find similarity amongst the complete records).

**Table 1**

Semantic similarity of attributes in the same query

| $q_1$ \ $q_1$ | $a_{11}$ | $a_{21}$ | $a_{31}$ | $a_{41}$ | $a_{51}$ |
|---|---|---|---|---|---|
| $a_{11}$ | - | 0.85 | 0.82 | 0.33 | 0.13 |
| $a_{21}$ | - | - | 0.91 | 0.14 | 0.65 |
| $a_{31}$ | - | - | - | 0.55 | 0.23 |
| $a_{41}$ | - | - | - | - | 0.22 |
| $a_{51}$ | - | - | - | - | - |

On the contrary to the existing methods [5], our system relies on the WordNet knowledge base, thus, the distance between the two concepts is always equal irrespective of their order ( i.e., $\delta(a_1,a_2) = \delta(a_2,a_1)$), because the ancestors of the attributes in a taxonomic branch are always common (as discussed in Equation (1)). Hence, we can avoid the repetition of such attributes in order to save computation time (as ignored in the existing methods). The distinct attributes that are obtained from each query record (e.g., query $q_1$= {$a_{11}$, $a_{41}$, $a_{51}$} retrieved in Example 2), can be used to compute the similarity score between the other query records. For this purpose, a set of 'n' queries are processed in non-repetitive ordered pairs, where the similarity score of the attributes is stored in a matrix. To do so, the similarity of each attribute $a_{mi}$ of a query record qi is measured with the attributes $a_{nj}$ of the other query $q_j$ (i.e., $q_i$ x $q_j$). This relation ($\Gamma(q_i,q_j)$) of matching queries is illustrated in Equation (3). As a result, several matrices are constructed that hold a similarity score of the attributes of the distinct pair of the queries (e.g., $q_i$, $q_j$) (as shown in Table 2). The similarity score of the queries is computed through Equation (2). This procedure is repeated in a non-repetitive order to avoid the repetition of the matching queries (as the similarity of queries ($q_i$, $q_j$) = ($q_j$, $q_i$)).

$$\Gamma\left(q_i,q_j\right) = \underset{m=1,k=1}{\overset{m=y,k=z}{sim}}\left(a_{mi},a_{nj}\right). \quad (3)$$

Hence, the total number of matrices to process '*n*' queries is determined from Equation (4). In this equation, the 2-permutations of *n* queries are the ordered arrangements of a query set represented in the matrices, which are arranged in a non-repetitive ordered pair. The constant factor '2' illustrates that each matrix must hold the comparison of two queries at a time. For example, we require three matrices to process three queries (n=3) (i.e., $q_1$, $q_2$, $q_3$), and the non-repetitive arrangement of these matrices will be ($q_1$ x $q_2$), ($q_1$ x $q_3$) and ($q_2$ x $q_3$). In these matrices, the paired attributes of the queries are also compared in a non-repetitive order to save computation time. Thus, we take the Cartesian product of each ordered pair of queries to compute the similarity between the attributes (which is required to determine the similarity of the records in the later phase).

$$\text{Matrices} = \frac{n!}{2*\left(n-2\right)!}. \quad (4)$$

**Definition 2:** A set of queries Q = {$q_1$,$q_2$, .....$q_n$}, where each query $q_n$ has attributes $q_n = \{a_{1n}, a_{2n}, .......a_{ln}\}$. Thus, a Cartesian product of the queries $q_i$ x $q_j$ has the following property (Equation (5)):

$$q_i \times q_j = \left\{(a_{mi},a_{nj}) \mid \left(a_{mi} \in q_i\right) \text{and} \left(a_{nj} \in q_j\right)\right\}, \quad (5)$$

where $a_{mi}$ symbolizes the m[th] attribute of the i[th] query, and a pair of attributes $a_{mi}$ & $a_{nj}$ (belong to queries $q_i$ and $q_j$ respectively) are examined mutually for semantic similarity.

For example, the following matrices (i.e., Table 2) illustrate the Cartesian product of three queries (i.e., $q_1$ x $q_2$, $q_1$ x $q_3$, and $q_2$ x $q_3$. The Cartesian product of the attributes of a query q1 (i.e., $a_{11}$, $a_{41}$, $a_{51}$) is computed with each attribute of query $q_2$ (i.e., $a_{12}$, $a_{22}$, $a_{32}$, $a_{42}$) (i.e.,

**Table 2**

The semantic similarity score of attributes of multiple queries

| $q_1$ \ $q_2$ | $a_{11}$ | $a_{41}$ | $a_{51}$ |
|---|---|---|---|
| $\mathbf{a_{12}}$ | 0.8 | 0.4 | 0.3 |
| $\mathbf{a_{22}}$ | 0.2 | 0.7 | 0.6 |
| $\mathbf{a_{32}}$ | 0.6 | 0.1 | 0.5 |
| $\mathbf{a_{42}}$ | 0.1 | 0.8 | 0.6 |

(a) $\Gamma(q_1,q_2)$

| $q_1$ \ $q_3$ | $a_{11}$ | $a_{41}$ | $a_{51}$ |
|---|---|---|---|
| $\mathbf{a_{13}}$ | 0.3 | 0.2 | 0.3 |
| $\mathbf{a_{23}}$ | 0.1 | 0.8 | 0.4 |
| $\mathbf{a_{33}}$ | 0.6 | 0.2 | 0.6 |
| $\mathbf{a_{43}}$ | 0.9 | 0.4 | 0.1 |

(b) $\Gamma(q_1,q_3)$

| $q_3$ \ $q_2$ | $a_{13}$ | $a_{23}$ | $a_{33}$ | $a_{43}$ |
|---|---|---|---|---|
| $\mathbf{a_{12}}$ | 0.6 | 0.2 | 0.8 | 0.4 |
| $\mathbf{a_{22}}$ | 0.5 | 0.2 | 0.7 | 0.5 |
| $\mathbf{a_{32}}$ | 0.5 | 0.3 | 0.7 | 0.3 |
| $\mathbf{a_{42}}$ | 0.4 | 0.6 | 0.1 | 0.6 |

(c) $\Gamma(q_2,q_3)$

Table 2(a)), and, in the same way, the similarity score is computed for other queries. In this table, the attributes having a similarity score greater than 0.8 (e.g., ($a_{11}$, $a_{12}$) =0.8) are more semantically similar than the other pair of the attributes. It is important to note that the distinct attributes of the queries are only equated in these matrices to conserve the computation time (as calculated in Example 2 of the semantic analysis).

The similarity between the attributes is computed through the following Algorithm 2. In this algorithm, the attributes of each query are semantically analyzed with the attributes of other queries, and their similarity score is computed using Equation (3) (lines 2-6). Moreover, the indices of semantically similar attributes (procedure detailed in Algorithm 1) of ordered paired queries are stored (lines 7-14), which are used to construct clusters in the later phase.

**Algorithm 2:** *SimilarityOfQueries* (set of *queries)*

1: set n ← total number of queries
2: **loop** starts at $h$ ← 1 till queries $q_n$-1 are processed
3: **loop** starts at $i$ ← $h$+1 till queries $q_n$ are processed
4: **loop** starts at $j$ ← 1 till total attributes of query $q_h$
5: **loop** start at $k$←1 till total attributes of query $q_i$
6: *Similarity[j][k]*=(1-*TaxonomicDistanc*
7: ($q_h$*_attributes[j],* $q_i$*_attributes[k]*))
8: **if** *similarity[j][k]* >= max
9: *indexes* ← Save indexes $h$ and $i$ of queries $q_h$
10: and $q_i$ respectively, and also the indexes of their matching attributes (i.e., $j$ and $k$)
11: **end if**
12: **end loop**
13: **end loop**
14: **end loop**
15: **end loop**
16: **return** *indexes*

As stated in Section 3, the proposed system relies on the MDAV algorithm that requires a distance of the records of a dataset in order to aggregate them in clusters. Therefore, we require a semantic distance of the query records in order to construct homogeneous clusters. Section 3.2 states the procedure to calculate the distance of the query records based on the similarity score of the attributes as calculated in this section.

## 3.2. Semantic Distance of the Records

As a result of semantic analysis (Section 3.1), we have a set of matrices for each ordered pair of the queries, in these matrices, each cell holds a similarity score of the matching attributes of the paired queries (as shown in Table 2). Since we require a semantic distance of the records in order to compute the centroid of the set-valued data (which is required to construct clusters in the later phase); therefore, it is necessary to develop a solution that measures the semantic distance of the multivariate of the set-valued records. The existing approaches [5, 28] calculate the weight of the attributes in terms of their recurrence/similarity within the same record, which is later used in determining the distance of a complete record (i.e., this weight factor of attributes is multiplied with the total distance of the record). This practice may affect the actual distance of the records, as a result, an incorrect centroid is chosen. Consequently, the redaction methods generate poor generalizations that cause huge information loss to the data. In contrast, our solution considers only one instance of semantically similar attributes of a query record while computing the distance of a complete record; however, the same generalization is enforced on all semantically similar attributes of the records during the redaction phase. In addition, we rely on semantic similarity scores of

**Figure 3**

Semantically similar and non-similar attributes of the queries the attributes in order to compute the distance of the records, hence, the semantic distance of the records is proportional to the number of semantically similar and non-similar attributes of the paired queries.

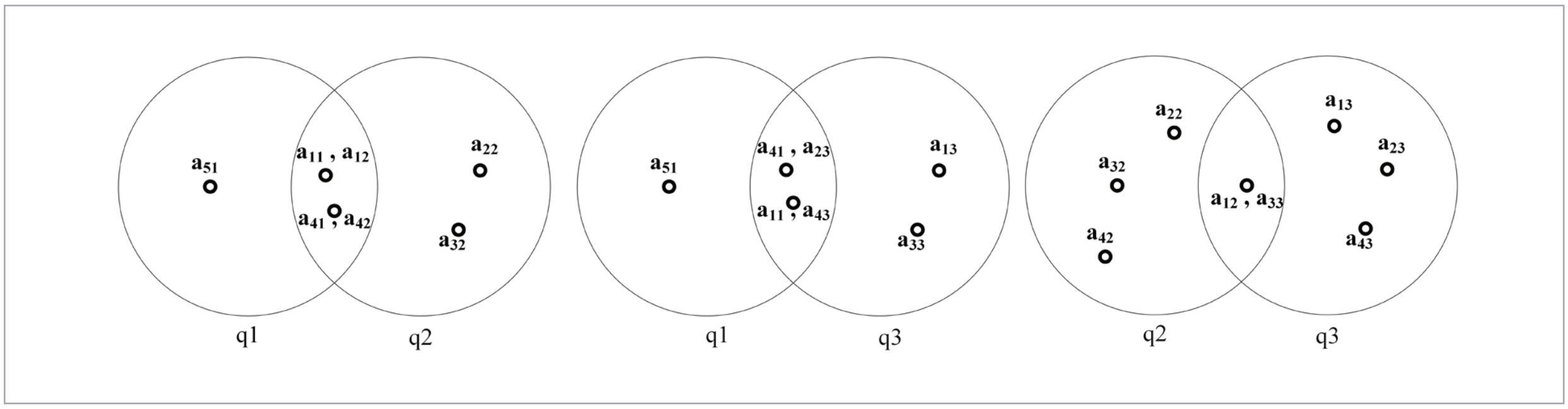

For this purpose, we require the cardinality of (i) semantically similar and (ii) non-similar attributes of the paired queries. As an example, Figure 3 illustrates semantically similar and non-similar attributes of the paired queries (i.e., $q_1$, $q_2$, and $q_3$) that are retrieved from the similarity scores as listed in Table 2. Hence, the cardinality score of semantically similar attributes of the paired queries can be computed through Equation (6). In this equation, it states that the cardinality of similar attributes of the paired queries (e.g., $q_i$ and $q_j$) is the difference between (i) the total number of attributes of the paired queries (i.e., $n(q_i)+n(q_j)$ ) and (ii) the distinct attributes that are indifferent in semantics (i.e., $n(q_i \cup q_j)$). The distinct attributes of the paired queries can be easily computed through the similarity table (i.e., Table 2) of the paired queries. In this equation (i.e., Equation (6)), only one instance of the paired attributes is considered that holds a similarity score greater than '0.8'. For example, in Table 2(a), the attributes of the queries $q_1$={ $a_{11}$, $a_{41}$, $a_{51}$}, and $q_2$={ $a_{12}$, $a_{22}$, $a_{32}$, $a_{42}$} have their cardinal values $n(q_1)$=3 and $n(q_2)$=4, whereas the indifferent attributes in these queries are $n(q_1 \cup q_2)=5$ (where three attributes $a_{51}$, $a_{22}$, and $a_{32}$ are distinct attributes and two pairs are semantically similar, i.e., ($a_{11}$, $a_{12}$) and ($a_{41}$, $a_{42}$) > 0.8). Hence, the cardinality of semantically similar attributes in these queries is $n(q_2 \cap q_3)=2$ .

$$n(q_i \cap q_j) = n(q_i) + n(q_j) - n(q_i \cup q_j). \qquad (6)$$

Similarly, the cardinality of non-similar attributes is computed by taking the symmetric difference of the paired queries (i.e., $q_i \Delta q_j$) (as stated in Equation (7)). The symmetric difference is the ratio between the distinct attributes ( $n(q_i \cup q_j)$ ) and the semantically similar attributes $n(q_i \cap q_j)$ (as computed in Equation (6)). For this purpose, the similarity score of the attributes of the matching queries can be inferred from Table 2. For example, the cardinality of distinct attributes of queries q1 and q2 is $n(q_1 \cup q_2)=5$ and the cardinality of semantically similar attributes is $n(q_i \cap q_j)=2$ (as computed in Equation (6)). Therefore, the symmetric difference of these queries is $n(q_1 \Delta q_2)$=2.5

$$n(q_i \ \Delta \ q_j) = \frac{n(q_i \cup q_j)}{n(q_i \cap q_j)}. \qquad (7)$$

The results retrieved through Equation (7) are used to compute the semantic distance of the paired queries (i.e., $\delta(q_i, q_j)$). The semantic distance of the corresponding queries is estimated to measure the degree to which the queries are semantically similar (i.e., the matching queries are less semantically similar for higher distance scores and vice versa). To achieve this, we drive the following Equation (8) that determines the distance of the queries. In this equation, the semantic distance between the paired queries (i.e., $\delta(q_i, q_j)$ ) is proportional to the ratio between (i) the number of non-semantically similar attributes (i.e., the symmetric difference of the paired queries (i.e., $n(q_i \Delta q_j)$) and (ii) their total number of distinct attributes (i.e., $n(q_i \cup q_j)$). For example, queries $q_1$ and $q_2$ in Figure 3 have 7 attributes in both queries. Therefore, the symmetric difference of the paired queries $q_1$ and $q_2$ is $n(q_1 \Delta q_2)$ =2.5, whereas, the value of the distinct attribute is $n(q_1 \cup q_2)=5$ (as retrieved from the trailing example of Equation (6) and Equation (7)). Therefore, the semantic distance of queries $q_1$ and $q_2$ is $\delta(q_1, q_2) = 0.5$ . Based on this, we can construct clusters that hold semantically coherent queries that have the least distance within a set of queries. To record such coherence between queries, we determine the distance of all possible non-repetitive ordered pairs of 'n' queries, and their results are stored in the distance matrix. We explain this procedure through the following Example 3.

$$\delta(q_i, q_j) = \frac{n(q_i \Delta q_j)}{n(q_i \cup q_j)}. \quad (8)$$

**Example 3:** Following Table 3 illustrates the distance of ten queries calculated through the stated procedure; these queries are collected from AOL query logs released in 2006 that are publically available. In this table, each tuple illustrates the distance of a distinct query with all possible combinations of other queries in non-repetitive ordered pairs. The distance between two queries ranges between '0' and '1' that signifies the level of variance of both queries (i.e., '0' specifies perfectly matched queries, and '1' signifies strongly dissimilar queries with all the attributes that differ from each other).

Once the semantic distance of all paired queries is determined, we need to compute the centroid of the query set to aggregate query records in their respective clusters, and then anonymize them accordingly. Following Section 3.3 elaborates these procedures.

## 3.3. Formation of Clusters and Anonymization

As the proposed system extends the MDAV algorithm, hence, the centroid of the dataset is required that is used to aggregate records in homogeneous clusters. Furthermore, these clusters are anonymized based on the centroid of their respective clusters. In the MDAV algorithm, it is trivial to compute the centroid of continuous datasets, however, the process to compute the centroid of the categorical dataset is a challenging task [28]. In addition, the imprecise selection of the centroid causes information loss during the data generalization phase. The state-of-the-art solutions [28] compute the centroid of the categorical datasets by driving the semantic distance of the multivariate of the records from the taxonomy. However, it is a computation-intensive solution to derive semantics of all multivariate records from the taxonomic database during each iteration of the centroid selection process. In contrast, we choose centroid of the set-valued data based on the cumulative distance of the entire query (instead of each attribute of the records). To do so, we compute the cardinality of the attributes of a record (as discussed in Section 3.2), which is sued to measures the degree of commonalities between a pair of the records (i.e., the cardinality of semantically similar and non-similar attributes).

To achieve this, Equation (9) measures the centroid of the dataset by relying on the semantic distance as computed in Table 3. In this equation, a set S holds such pairs of queries that have minimum distance amongst their respective tuples of the distance table (collected from the tuples of Table 3). A pair with the minimum distance constitute that the query qi shares maximum semantically similar attributes with the other pair of the query (e.g., $q_j$) (in contrast to the rest of the queries). Therefore, we collect such pairs of queries in clusters that have minimum distance amongst the other pairs to increase the cohesiveness of clusters. In addition, it is mandatory to choose a pair of queries from each tuple to guarantee the presence of each query in the final clusters (regardless the matching query has been chosen in other tuples). Then, the centroid distance is computed by taking the mean of the distance score of all pairs of a set S (where N is the total number of pairs in this set).

**Table 3**
Distance of distinct pair of queries

| | $q_1$ | $q_2$ | $q_3$ | $q_4$ | $q_5$ | $q_6$ | $q_7$ | $q_8$ | $q_9$ | $q_{10}$ |
|---|---|---|---|---|---|---|---|---|---|---|
| **$q_1$** | - | 0.50 | 0.33 | 0.88 | 1.0 | 0.88 | 0.88 | 0.11 | 0.88 | 0.75 |
| **$q_2$** | - | - | 0.40 | 0.75 | 0.88 | 1.0 | 0.88 | 0.57 | 0.57 | 0.75 |
| **$q_3$** | - | - | - | 0.75 | 1.0 | 0.75 | 0.88 | 0.38 | 0.88 | 0.88 |
| **$q_4$** | - | - | - | - | 0.57 | 1.0 | 0.88 | 0.88 | 0.88 | 0.57 |
| **$q_5$** | - | - | - | - | - | 0.75 | 0.75 | 1.0 | 0.57 | 0.75 |
| **$q_6$** | - | - | - | - | - | - | 0.75 | 0.88 | 0.88 | 0.75 |
| **$q_7$** | - | - | - | - | - | - | - | 0.75 | 0.75 | 1.0 |
| **$q_8$** | - | - | - | - | - | - | - | - | 0.88 | 0.75 |
| **$q_9$** | - | - | - | - | - | - | - | - | - | 0.75 |
| **$q_{10}$** | - | - | - | - | - | - | - | - | - | - |

$$\left.\begin{aligned} S &= \arg\min_{q_i, q_j \in Q} \left\{ \sum_{i=0}^{n-1} \overset{n}{\underset{j=i+1}{\delta}} (q_i, q_j) \right\} \\ centroid_score &= \frac{\sum S}{N} \end{aligned}\right\} . \tag{9}$$

The empty cells in the table (marked with a dash (-) as shown in Table 3) do not affect the overall results, because the distance of such paired queries is computed and analyzed in the other cells of the tuples (e.g., the distance of a pair ($q_4$, $q_2$) is same as computed for ($q_2$, $q_4$)). Finally, the centroid pair of the queries from a set S is chosen that has the semantic distance closer to the centroid score of the set S. In case there are more than one pairs that have a similar distance score, then we can choose any pair as the centroid pair because these pairs will share a common cluster as they have an equal number of semantically similar attributes. This process is explained through the following Example 4 (which is the extension of Example 3).

**Example 4:** The following Table 4 shows the optimal query pairs and their respective distance scores that are computed through Equation (9). This table is the illustration of a set S that holds the least distant pairs determined from each tuple of Table 3. Table 4 holds pairs for each distinct query in order to guarantee the presence of each query in the final query set that will be used to construct clusters in the later phase. We choose a minimum distance score from each tuple of the distance table (as highlighted in Table 4). The queries are compared in non-repetitive ordered pairs. For example, in the pairs for query $q_2$, the following pair of queries ($q_2$, $q_3$) have the minimum distance score (i.e., 0.40) from the rest of the pairs, therefore, this pair is chosen for onward steps. Moreover, we ignore other pairs that are marked with dash notation (-) because such pairs have already been compared in the previous tuples of the table. For example, in a tuple for query $q_8$, the minimum distance pair for this query is $q_1$, but this pair ($q_1$, $q_8$) has already been compared in the previous tuple for query $q_1$. Moreover, if we repeatedly take the least distant pairs then we may ignore some of the queries in the clustering phase; as a result, the missing queries will not be anonymized in the later phase.

Based on Table 4, the centroid score computed for this table is 0.59. Therefore, we choose a query pair as the centroid that has a distance score closer to the centroid score. Preferably, a query pair that has a distance smaller than the centroid score is chosen because it holds a more semantically similar multivariate in contrast to the pairs that have a larger distance score. In this case, we have four pairs that have smaller distance scores (i.e., 0.57), which are highlighted in Table 4. Hence, we can choose any pair as the centroid because all of these pairs will share a common cluster as they have an equal number of multivariate that are semantically similar. The rest of the procedure to compute clusters and to anonymize data is discussed in the following algorithm.

### 3.3.1. Adaptive-MDAV Algorithm

As mentioned in Section 1.1, our method relies on an adaptive size clustering approach in which the records are aggregated in clusters based on their semantics. However, the minimum size of the clusters is adhered in order to comply with the notion of k-anonymity principle. This type of approach is more cohesive [28] as it accumulates set-valued records based on their common properties (i.e., the semantic similarity of the attributes). As a result, the clusters emit less information loss during the data generalization phase. Hence, we incorporate the self-adaptiveness into the MDAV algorithm by clustering only those pairs of queries that have similar/contiguous distances. The similarity in the distances of the query pairs indicates that these pairs hold an equal number of semantically similar attributes. To achieve this, the following modifications are made to the existing MDAV algorithm: (i) an adaptive size clusters, and (ii) the query records are chosen for cluster based on their maximum number of semantically similar multivariate. The modified MDAV algorithm is illustrated below (i.e., Algorithm 3) that highlights the processes of cluster formation and the anonymization of the records of these clusters.

**Table 4**
Least Distance of Paired Queries

| $q_1,q_8$ | $q_2,q_3$ | $q_3,q_8$ | $q_4,q_{10}$ | $q_4,q_5$ | $q_5,q_{10}$ | $q_5,q_9$ | $q_6,q_{10}$ | $q_6,q_7$ | $q_7,q_8$ | $q_7,q_9$ | $q_8,q_{10}$ | $q_9,q_{10}$ |
|---|---|---|---|---|---|---|---|---|---|---|---|---|
| 0.11 | 0.40 | 0.38 | 0.57 | 0.57 | 0.57 | 0.57 | 0.75 | 0.75 | 0.75 | 0.75 | 0.75 | 0.75 |

This Algorithm 3 operates on a set of paired queries $Q_p$ along with their paired distances D that illustrates the amount of semantically similar multivariate of these pairs (as calculated above in Table 4) (lines 1-3). It is important to note that such pairs are optimal in terms of semantic similarities, which are retrieved after the semantic analysis of the query records. This algorithm aggregates at least k query records in each cluster to comply with the principle of k-anonymity, however, the size of the clusters is not fixed (i.e., clusters size >= k) (line 4). In addition, the query logs that are smaller than the size k are ignored and such records are never published. In order to construct clusters, the centroid of the dataset (i.e., Table 4) is computed (line 5) (details in Section 3.3). Based on this centroid, the following two pairs of the queries are marked: (i) a pair that is most distant to the centroid (i.e., $(q_i,q_j)$), and (ii) a pair that is most distant to a pair computed in step (i) (i.e., $(q_x,q_y)$) (lines 6-7). In contrast to the conventional MDAV algorithm, we do not fix the size of the clusters but the clusters are adaptive in nature. To achieve this, the pairs of queries that are closer to such pairs (i.e., in lines 6 & 7) are aggregated in their respective clusters (lines 8 &10). It is important to note that the distance score of a pair $(q_x,q_y)$ is smaller than the pair $(q_i,q_j)$. A pair with a smaller distance score holds maximum semantically similar attributes, therefore, the cluster for this pair (i.e., $(q_x, q_y)$) forms a cluster before the other pair (i.e., $(q_i,q_j)$). The clusters hold only distinct instances of the paired queries (e.g., cluster $c_1=\{q_1,q_2,q_3,q_4\}$). In addition, all remaining pairs in a set $Q_p$ that contains any of the instances of the clustered queries are removed from the set (lines 8 & 11). Similarly, the above-mentioned procedure is repeated for

**Algorithm 3:** *Adaptive_MDAV* (*paired_queries, istance_of_paired_queries*)

1: $Q_p$=*paird_queries*
2: D=*distance_of_paired_queries*
3: $Q^A$ (clusters of queries)
4: **while** ($Q_p$ has paired queries > *k*)
5: Calculate centroid μ of the paired queries of $Q_p$
6: Find the most distant paired query $q_i$, $q_j$ to the centroid μ
7: Find the most distant paired query $q_x$, $q_y$ to the queries $q_i$, $q_j$
8: Construct a cluster in $Q^A$ with queries $q_x$, $q_y$, and all records that have distance scores closer to the queries $q_x$, $q_y$
9: Remove all records from query sets $Q_p$ that has any element of these paired queries i.e., $q_x$ or $q_y$
10: Construct a cluster in $Q^A$ with queries $q_i$, $q_j$, and all records closest to the queries $q_i$, $q_j$
11: Remove these records from query sets $Q_p$ that has any element of these queries i.e., $q_i$ or $q_j$
12: **end while**
13: Construct a cluster in $Q^A$ with the remaining records
14: **for** each cluster in $Q^A$ do
15: Compute the centroid of the cluster
16: Replace all semantically similar attributes of the records with the centroid values
17: Replace non-similar attributes of a record with the centroid attributes of the taxonomic branch
18: **end for**

the remaining pairs in a set $Q_p$ until all queries are aggregated in their respective clusters.

At this point, we have several clusters and each cluster holds a set of distinct homogeneous query records. Now, we can anonymize the query records of each cluster with the help of their respective centroids. Hence, it is essential to compute the centroid record of each cluster. To compute a centroid (line 15), we choose a record from each cluster that has maximum cardinality in terms of recurrence in the pairs that are clustered together, and in case of a tie, we also consider the least paired score to break the tie. The maximum cardinality of a record illustrates that it has maximum semantically similar attributes to the other records. In case of conflict in the recurrence of the queries, any query record can be chosen in order to break the tie. Now, we can anonymize query records of the cluster by replacing them with the centroid record to make them indistinguishable (as explained in Section 1). As query logs are unstructured in nature, therefore, these records may not be perfectly semantically similar (i.e., the number of semantically similar and non-similar attributes may vary). As a result, if we replace all attributes of the records with the attributes of the centroid record (in order to anonymize them); it may result in huge information loss, and such records may not be useful for analytical purposes. To deal with this situation, we replace only those attributes of the query record that are semantically similar to the attributes of the centroid record. Whereas, the rest of the data items (that are not semantically similar to any other attribute of the centroid record) are generalized with the centroid node of the taxonomic branch of the respective data items (as mentioned in Figure 2) (lines 14-18). The taxonomic branch of the data item holds all semantically similar data values, therefore, it preserves the semantics of the data item. This approach retains the actual semantics of the attributes to minimize information loss. The working of this algorithm is explained in the following Example 5.

**Example 5:** We extend Table 4 to demonstrate the working of Algorithm 3. The following results are computed:

Centroid score of all paired records = 0.59

A pair of queries close to the centroid is $\mu = (q_4, q_5)$ (i.e., score =0.57)

Distant pair to $\mu$ is $(q_9, q_{10})$

Distant pair to queries $(q_9, q_{10})$ is $(q_1, q_8)$

Cluster $c_1$ holds all query pairs that are closer to the distance score of a pair $(q_1, q_8)$, which are $(q_2, q_3)$ and $(q_3, q_8)$, hence, the cluster $c_1=\{q_1, q_2, q_3, q_8\}$. Similarly, the other clusters are $c_2=\{q_6, q_7, q_9, q_{10}\}$ and $c_3=\{q_4, q_5\}$.

Based on this, a record from each cluster is chosen as a centroid that has maximum cardinality in terms of recurrence; in addition, we choose the least paired score to break the tie (as explained above). As a result, the centroids $(\mu_n)$ of the respective clusters are $\mu_1= q_8$, $\mu_2= q_7$, and $\mu_3= q_5$. These centroids can replace the query records of their respective clusters as per the rules defined in Algorithm 3.

## 4. Evaluation

In this section, the proposed system is evaluated to measure its significance in terms of the homogeneity of the clusters, and the utility of the anonymized records. For this purpose, the proposed system is evaluated based on the following two aspects, which are (i) cohesion of the clusters (i.e., in terms of homogeneous records), and (ii) utility of the anonymized data. First, we explain the measures used for the evaluation of the proposed mechanism (Section 4.1). Then, in Section 4.2, we detail the analysis of our proposed system based on these measures.

### 4.1. Evaluation Measures

As introduced in Section 1, the records are aggregated in clusters based on their semantic similarity. In addition, these records are anonymized by preserving their semantics during the microaggregation process. Hence, we require such metrics that could gauge the cohesion of clusters and the information loss of the records generated as a result of microaggregation methods. Therefore, we derived two methods to gauge the significance of the proposed method: (i) cohesion of clusters and (ii) information loss of clusters. The cohesion of a cluster illustrates the amount of semantic dispersion of the records sharing a common cluster, whereas, the information loss is measured to determine the utility of data generated as a result of anonymization methods.

In order to measure the cohesion of clusters $(\check{C}_c)$, we propose a measure that computes the degree of dispersion of the records from the centroid of the cluster (i.e., semantics dispersion of the records). We drive a metric to measure the cohesion of clusters from the existing literature [1, 10]. The existing metric (i.e., Sum square of distances SSE) computes the distance between the attributes of the records and the centroid attributes of a cluster in order to measure the homogeneity. In view of this metric, we measure the homogeneity of the query records with respect to the centroid of the cluster. For this purpose, we drive the semantic distance of the queries and the centroid of the clusters to gauge the dispersion of records from the centroid. To achieve this, the proposed measure takes account of the semantically similar and non-similar attributes of the examined records. Hence, the following Equation (10) computes the degree of dispersion of the query records from the respective centroid of the clusters. This equation states that the cohesion of a cluster $\zeta_k$ (that holds n set of queries, i.e., $\zeta_k = \{q_1, q_2, ....q_n\}$) is measured through the sum of semantic variations between the query records of a cluster $\zeta_k$ and its centroid $\alpha_k$. For this purpose, we use a distance measure defined in Equation (8) (i.e., $\delta(q_i, \alpha_k)$ to compute the sum of semantic variation between the query records and the centroid $\alpha_k$, where n-1 is the total number of queries within a cluster $\zeta_k$ (excluding centroid of the cluster). This distance measure (i.e., Equation (8)) computes the ratio between the (i) non-similar attributes of the paired records (i.e., between query qi and the centroid of the cluster $\alpha_k$), and (ii) the total number of attributes of both records. In addition, a query record $q_j$ is chosen as a centroid of the cluster (i.e., $\alpha_k = q_j$) that has maximum cardinality in terms of semantically similar attributes to the other records of the cluster $\zeta_k$ (as detailed in Section 3.3). It is important to note that a single instance of all semantically similar attributes (within a query record) is considered while measuring the cohesiveness of the clusters, because multiple instances of these attributes in a query may disrupt the results. Therefore, we consider only one instance of these attributes in query $q_i$ when comparing with the other query $q_c$, since these two attributes used for the same purpose may increase/decrease the similarity score in relation to the other query. For example, a query $q_i$ has five attributes $q_i$={*peach, apple, orange, citrus, cherry*} and a query $q_j$ has three attributes $q_c$={*circle, orange, soccer*}. The attributes orange and citrus are semantically similar within a query $q_i$, and both attributes are semantically similar to orange in query $q_c$, hence, the mutual distance score for both of these queries will be low (i.e., high semantically similar). Hence, a cluster that has a low cohesive factor (which ranges between 0 to 1) indicates that the records it holds are closer to the centroid of the cluster, hence a cluster tends to be more cohesive.

$$\left.\begin{aligned} \check{C}c_{\zeta_k} &= \sqrt{\frac{1}{n-1}\sum_{i=1}^{n-1}\delta(q_i,\alpha_k)} \\ \because \alpha_k &= \arg\max_{q_j\in\zeta_k} card\left(q_j\right) \end{aligned}\right\} \quad (10)$$

In addition, the utility of anonymized data is essential for analytical purposes. Moreover, it is desirable that the data generated, as a result of microaggregation-based techniques, must illustrate the same semantics as the original data. The utility of data is attributed to the information loss (IL) that quantifies the change in the semantics of the original data and the anonymized data. Thus, in order to measure information loss (IL) of the anonymized dataset, we adapt a measure (i.e., Equation (11)) that is widely used by many researchers [1, 12, 28] for the said purpose. The information loss (measured in percentage) is computed as a ratio between the sum of the squared errors (SSE) of the clusters and the total sum of squares (SST) of the complete dataset, which is explained in the subsequent paragraphs.

$$IL = \frac{SSE}{SST}\times 100. \quad (11)$$

As already mentioned, the microaggregation-based techniques focus to aggregate homogeneous records in clusters and then anonymizing such records with the help of the centroid of their respective clusters. Therefore, in order to gauge the value of anonymized data, we compute the information loss of the anonymized records of the cluster generated through the anonymization method. For this purpose, we compute SSE that measures the amount of change in the semantics of the records that are occurred due to the anonymization process. As explained in Section 3.3, the semantically similar attributes of the records of a cluster are anonymized through the centroid record, and the attributes that are not semantically similar to any other attributes of the centroid record are replaced with the centroid node of the taxonomic branch retrieved from the ontological knowledge base (i.e., WordNet). Therefore, in order to compute SSE for the proposed system, we determine a semantic change in each record for both of these scenarios (i.e., the attributes that are semantically similar and that are not similar to the centroid record). This change can be measured by computing the semantic distance between the original data and the anonymized data (i.e., the distance between the original value of the attributes and the centroid values). Thus, we compute SSE through the following Equation (12). In this equation, the sum of the square of distances between each query qi and the centroid record $\alpha_k$ of the cluster or the taxonomic branch is measured as follows.

$$\underset{x_{mi} \in q_i, \mu_k \in \alpha_k}{SSE} = \sum_{i=1}^{n-1}\left[\sum_{m=1}^{j}\left(\delta\left(x_{mi}, \mu_k\right)^2 * w_m\right)\right]. \quad (12)$$

For this purpose, the sum of semantic distance is computed (through the measure defined in Equation (1) i.e., $\delta(x_{mi}, \mu_k)$) between (i) each attribute '$x_{mi}$' of a query $q_i$ (where $x_{mi}$ is the $m^{th}$ attribute of $i^{th}$ query) and (ii) it's corresponding semantically similar attribute $\mu_k$ of the centroid record or the attribute in the taxonomy. Whereas, the factor $w_m$ is the weight of the $m^{th}$ attribute that has multiple instances in a query $q_i$ (in terms of semantic similarity). In order to measure the distances of the attributes, we rely on Table 2 that holds the distance scores of the attributes of the paired records. Hence, the SSE holds the sum of the distance of all queries (which range between 1 to *n-1* excluding the centroid record) of the $k^{th}$ cluster.

Likewise, the total sum of squares (SST) determines the sum of the square of the distances between the individual records and the centroid of the overall dataset. In order to compute the centroid of an overall dataset, we rely on a measure (as defined in Equation (9)) that determines a paired query that has the least distance from the other pairs of the queries. Hence, any query $q_i$ of the chosen pair is considered as the centroid query ć (i.e., ć = $q_i$) of an overall dataset. The Equation (13) illustrates this measure to compute the SST, in which, the sum of the square of the distance between each attribute '$x_{mi}$' of a query $q_j$ and it's corresponding semantically similar attribute $\lambda_k$ of the centroid record is measured ($w_m$ is the weight of mth attribute). Similarly, the cumulative sum of all queries is measured, where queries range between 1 to *n - 1* excluding the centroid record.

$$\underset{x_{mi} \in q_i, \lambda_k \in ć}{SST} = \sum_{i=1}^{n-1}\left[\sum_{m=1}^{j}\left(\delta\left(x_{mi}, \lambda_k\right)^2 * w_m\right)\right]. \quad (13)$$

## 4.2. Evaluation Results

To measure the performance of our system, in this section, we compute (i) the complexity of our algorithms, (ii) the cohesiveness of the clusters, and (iii) the information loss comparison with respect to existing solutions.

First, we compute the complexity of Algorithm 1 that determines the similarity of the attributes of any query qi. For this purpose, the n attributes of a query qi are matched mutually to measure the similarity scores of the attributes. Therefore, the n attributes are matched with the rest of *n-1* attributes. Hence, the complexity of the algorithm is O(n*(n-1)). In Algorithm 2, we compute the similarity of the queries according to their matching attributes. Hence, each query is examined for the similarity with the rest of the queries. Therefore, the n queries are matched with the rest of *n - 1* queries and the computation cost is O(n*(n-1)). In addition, the n attributes of a query $q_i$ are matched with the n attributes of other query $q_j$. Hence, the computation cost of matching attributes is O(n2). Algorithm 3 creates the clusters of the paired queries retrieved as a result of Algorithm 2. In this algorithm, the n pair of queries are processed and matched with the other pairs based on their similarity score. Hence, the complexity of this algorithm is O(n).

The overall complexity of our method to match attributes and then generate clusters is $O(n^2+n)$. In comparison, the complexity of MDAV-based solution [38] to generate clusters is $O(n^2)$, whereas the complexity to generate clusters in another scheme [6] is $O(n^2+n/k)$. This comparison states that the complexity to generate the clusters in our approach is not affected by the proposed methods, and it is the same as other MDAV-based solutions [6, 38]. However, we have improved the information loss generated as a result of these methods that preserve the utility of the anonymized data (as shown in Figure 5).

We implemented our system in Java programming language with the supporting APIs (i.e., Stanford

**Figure 4**
The cohesiveness of Clusters

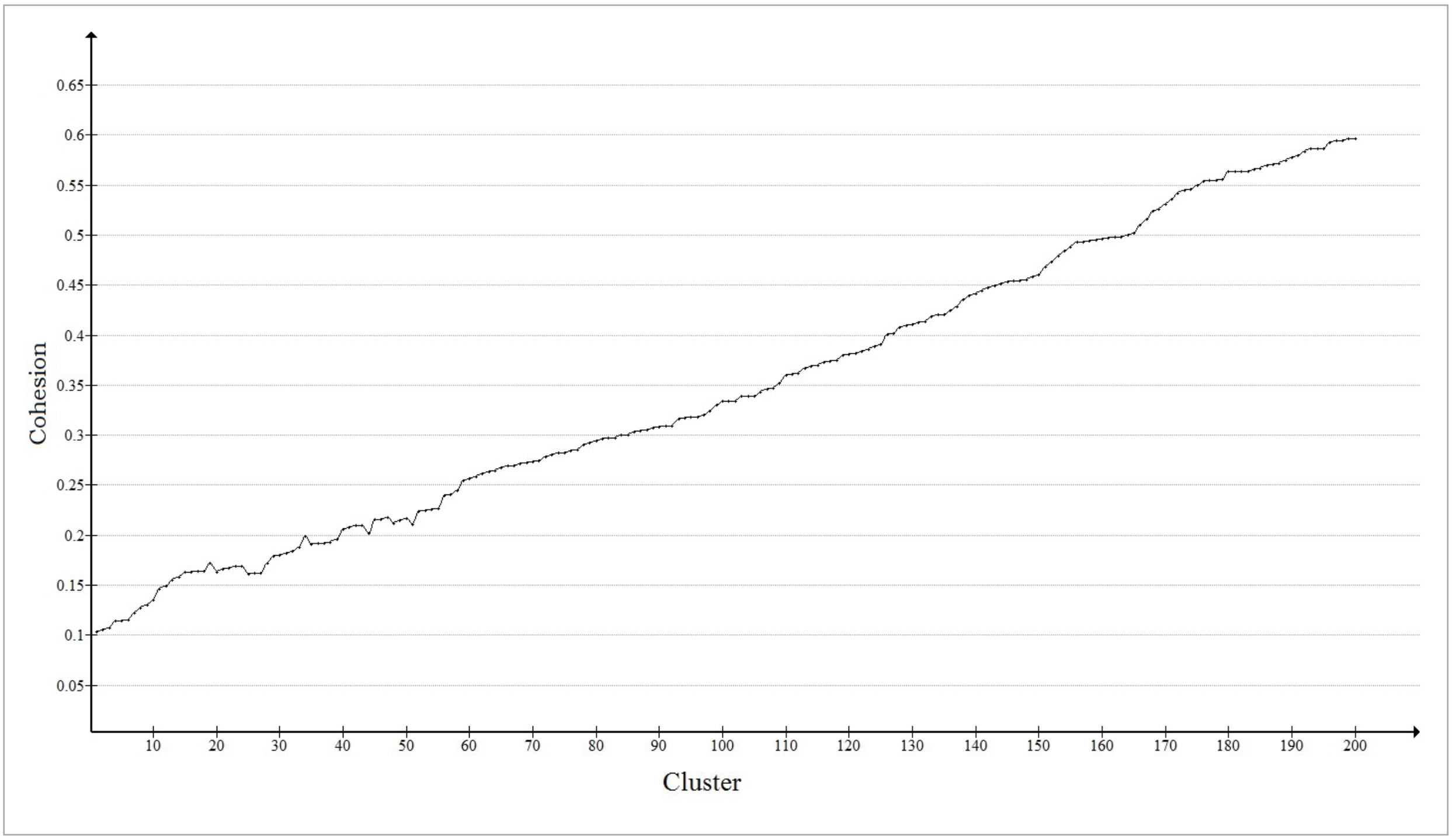

NLP[1], OpenNLP[2] , and Jena API[3]) that are used for the semantic analysis of the query logs through the WordNet ontology. In addition, we obtained a query log dataset that was published by AOL in 2006 (which is publically available at [3]). For semantic analysis, we obtained logs of 800 users and chose their 10,000 query records at random. Our system aggregated query records in 200 clusters, where each cluster holds variable-size query logs. The maximum size of a cluster that is observed during the experiments is 1600 queries and the minimum size of the cluster was 14 queries, whereas, the $k$ value set for these query logs was 5 queries. The proposed model complies with the minimum size of the cluster as defined in the algorithm (i.e., $k$ records); however, the size of the cluster is never fixed.

First, we measure the cohesiveness of the clusters in order to quantify the homogeneity of the records. For this purpose, we measure the degree of dispersion of the query records from their respective centroids by using Equation (10). As the measure to compute cohesiveness of the clusters (i.e., Equation (10)) relies on the distance method defined in Equation (8), thus, the difference between the query records of a cluster varies between '0' to '1'. Where, '0' implies that the query records of a cluster are convergent to the centroid; hence, we deduce that the cluster is more cohesive. Whereas, '1' states that all attributes of the records are semantically dissimilar and the cluster is non-cohesive. The results are illustrated in Figure 4 that shows the cohesion score of 200 clusters. From these results (i.e., Figure 4), it is important to note that 75% of the clusters have cohesion scores less than 0.45, which indicates that the query records that are aggregated in these clusters are more semantically similar. Moreover, there are few clusters (i.e., between 150-200) that have cohesive factor higher than 0.5, which implies that these clusters hold such query records that are least semantically similar to the former set of the clusters. Because such clusters hold leftover query logs (that do not perfectly match with other records), hence, these clusters have high scores of cohesiveness.

[1] https://nlp.stanford.edu/

[2] https://opennlp.apache.org/

[3] https://jena.apache.org/

In addition, we determined the utility of the anonymized data by computing the information loss caused by our system (as detailed in Section 4.1). Moreover, we compare the results of our work with the following state of art solutions that anonymize the categorical unstructured data:

Senavirathne et al. [32]: A machine learning-based anonymization approach that relies on probabilistic k-anonymity principal to anonymize set-valued data.

Majeed and Lee [27]: An anonymization method that relies on the machine learning approach to preserve the privacy of the published data, and then anonymizes the data by using generalization technique.

Batet et al. [5]: An approach that relies on an open directory project (ODP) to extract the semantics of the query logs, and then microaggregate such logs based on the derived semantics.

Novarro-Arribas et al. [31]: An anonymization method based on the syntactical measures that were proposed to compare the query logs.

Figure 5 illustrates the comparison of the proposed system and the existing solutions. Our solution is adaptive in nature (i.e., the size of the clusters is determined automatically after the semantic analysis of the set-valued data), hence, we collected different sizes of clusters that are automatically generated by the system. However, in this comparison, we choose only those clusters from the proposed solution that have the same size as in the existing solutions. In this analysis, we observed that our proposed model has similar results as Majeed et al. [27] for the cluster size ranges between 1 to 300, however, our model outperforms for the large size clusters. It can be seen through the graph that the information loss increases as the cluster size increases. In addition, there is a rapid increase in information loss for the cluster having a cluster size greater than 1400 records, however, the proposed scheme has less effect on the information loss as compared to other methods. Because such clusters hold a bulk of those records that are more semantically similar. Hence, we can conclude that our system perfectly aggregates semantically similar records in common clusters, and the information loss, caused as a result of the redaction method, is also

**Figure 5**

Information Loss (IL) Comparison

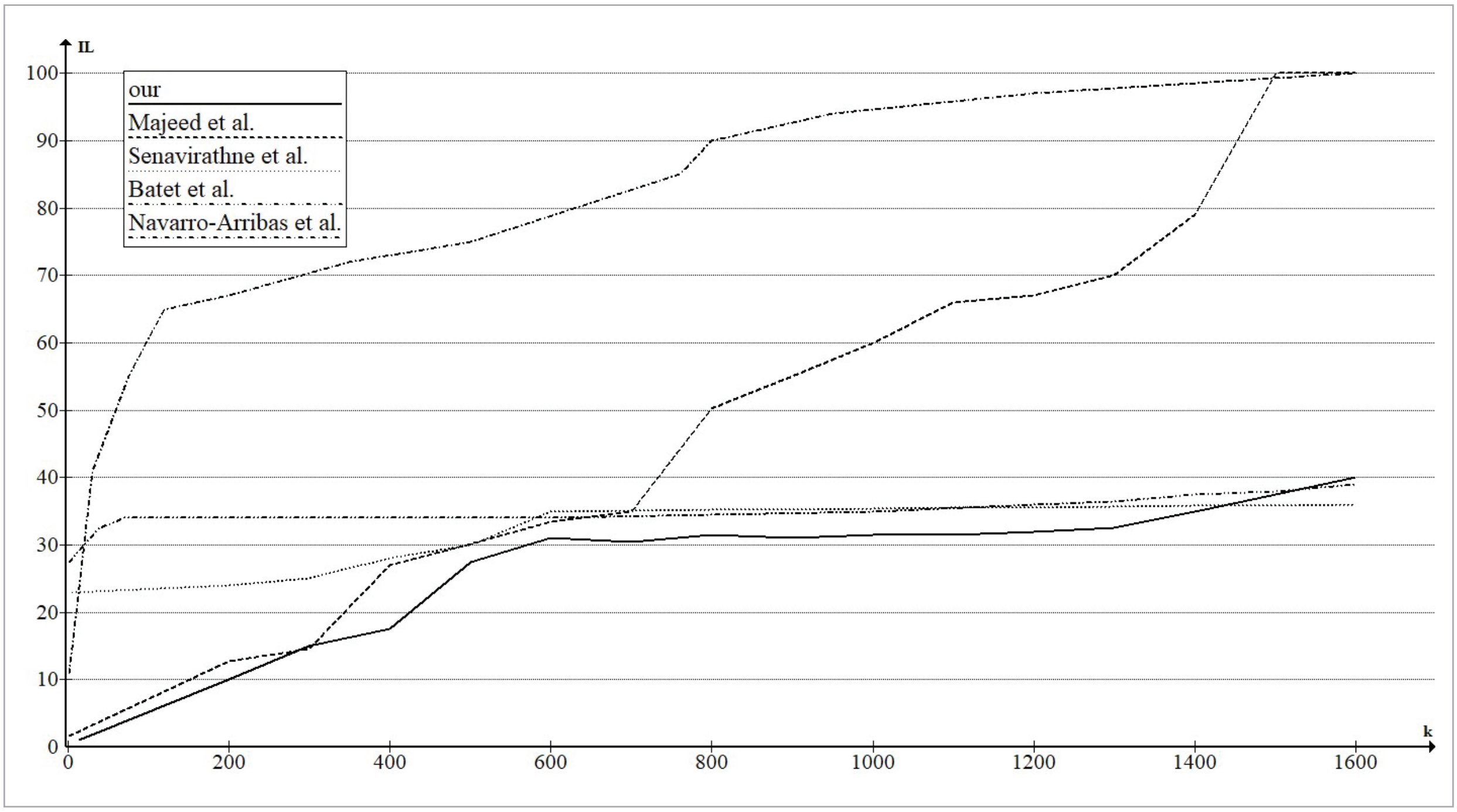

**Figure 6**
Information Loss in Random Clusters

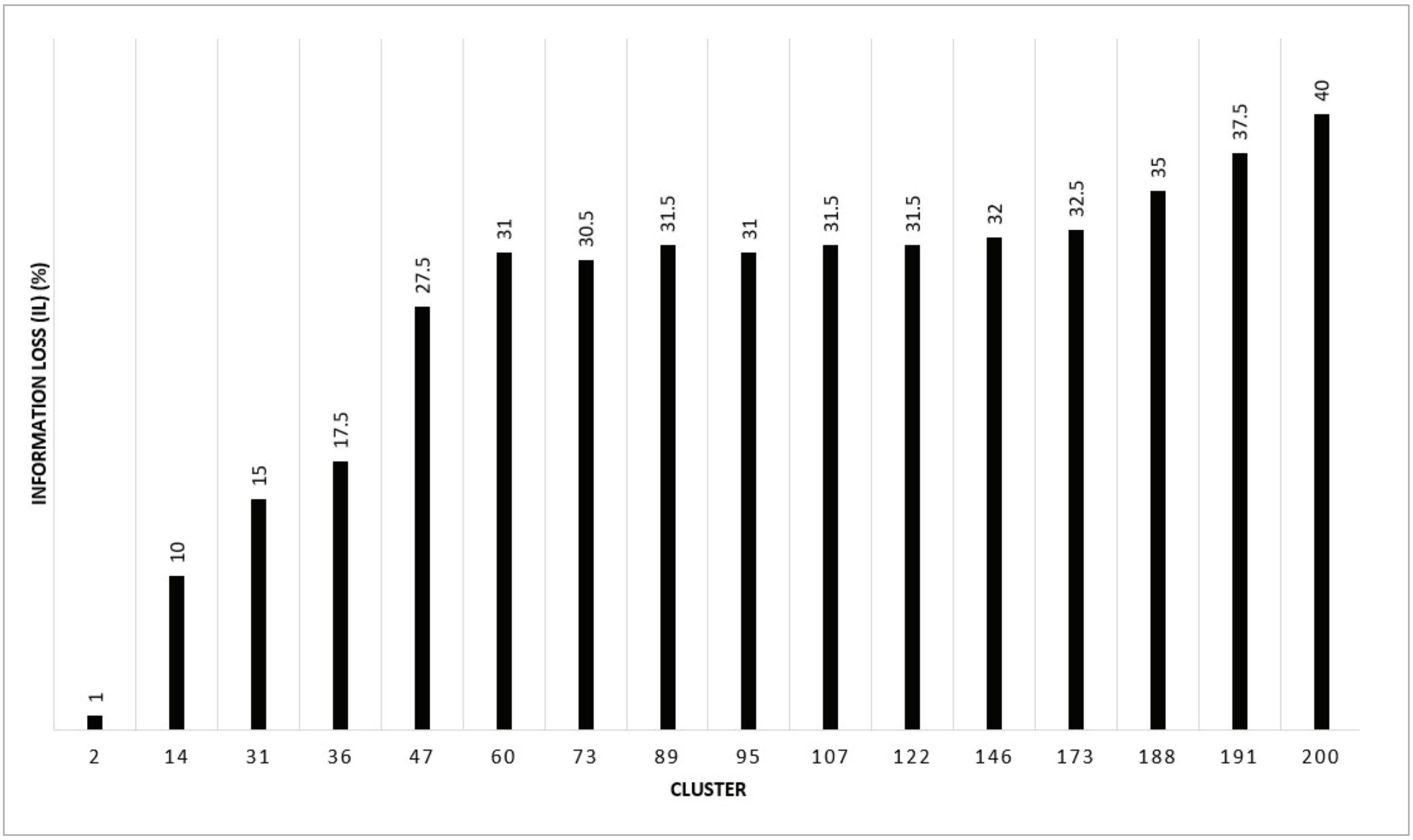

improved than the existing state-of-the-art solutions. In addition, the fixed-size clusters (i.e., existing solutions) are constrained to aggregate such records that may not be semantically coherent to other records, as a result, these solutions uphold more information loss caused by the redaction methods. However, our adaptive size clusters only aggregate those records that are semantically coherent to other records of the clusters, hence, the results are much improved.

In addition, we computed the information loss of random clusters that is illustrated in Figure 6 The stats show that the data loss is low for the first 40 clusters but it increases logarithmically for the rest of the clusters. However, the logarithmic increase in IL is minimal due to the homogeneity of the clusters. As shown in Figure 4 and Figure 5, the clusters formed in the beginning are more cohesive as there is less information loss. These clusters are independent of the cluster size but hold such query logs that are more semantically similar. Hence, the proposed MDAV algorithm is not affected by the size of the clusters.

## 5. Conclusion and Future Work

In this paper, we present a novel method to microaggregate set-valued data that aggregate semantically similar records holding sensitive information (i.e, identifying variables and sensitive variables). For this purpose, we rely on a taxonomic database (i.e., WordNet) to derive the semantics of the data items. To gauge the distance of the records, it measures semantically similar and non-similar multivariate of the records, which help to aggregate semantically similar records in common clusters. In addition, this method relies on an adaptive size clustering approach that accumulates records according to their semantics instead of fixed-size clusters. As a result, the clusters hold homogeneous records and the anonymized records generated as a result of redaction methods emit less information loss.

As future work, we plan to further improve the redaction method (detailed in the last paragraph of Section 3.3) by dealing with non-semantically similar data items of the records. These data items are only generalized to hide the details but they do not comply with the principle of k-anonymity (i.e., at least k-similar records to anonymize records). Hence, we plan to extend this solution with the partial clusters holding semantically similar and non-similar data items of the records. The partial clusters of non-semantically similar data items are grouped with the data items of other records. In addition, as this proposed work only relies on sensitive attributes, we plan to extend this work to microaggregate other types of variables from the unstructured data (i.e., quasi-identifiers). In addition, we plan to improve this mechanism to work with the differential privacy method to address the limitation of the k-anonymity model.

### Acknowledgement

We acknowledge Foundation University Islamabad for its support to conduct this research work.